# X-ray absorption spectroscopy (XAS) investigation of the electronic structure of superconducting FeSe$_x$ single crystals

C. L. Chen,[1]* S. M. Rao,[1]† C. L. Dong,[2] J. L. Chen,[3,4] T. W. Huang,[1] B. H. Mok,[1] M. C. Ling,[1] W. C. Wang,[3] C. L. Chang,[3] T. S. Chan,[2] J. F. Lee,[2] J.-H. Guo,[4] and M. K. Wu[1]

[1.] Institute of Physics, Academia Sinica, Taipei 11529, Taiwan
[2.] National Synchrotron Radiation Research Center, Hsinchu 30076, Taiwan
[3.] Department of Physics, Tamkang University, Tamsui, Taipei County, Taiwan
[4.] Advanced Light Source, Lawrence Berkeley National Laboratory, Berkeley, CA 94720

E-mail: *clchen@phys.sinica.edu.tw; † rao@phys.sinica.edu.tw

X-ray absorption spectroscopy (XAS) Fe K-edge spectra of the FeSe$_x$ (x=1-0.8) single crystals cleaved *in situ* in vacuum reveal characteristic Fe 4$sp$ states, a lattice distortion and the Se K-edge spectra point to a strong Fe 3$d$-Se 4$p$ hybridization giving rise to itinerant charge carriers. A formal charge of ~1.8+ for Fe and ~2.2- for Se were evaluated from these spectra in the FeSe$_x$ (x=0.88). The charge balance between Fe and Se is assigned itinerant electrons located in the Fe-Se hybridization bond. As x decreases the 4$p$ hole count increases and a crystal structure distortion is observed that in turn causes the Fe separation in the *ab* plane change from 4$p$ orbital to varying (modulating) coordination. Powder x-ray diffraction (XRD) measurements also show a slight increase in lattice parameters as x decreases (increasing Se deficiency).

## II. INTRODUCTION

Recent report of superconductivity in oxypnictides has generated a flurry of activity much akin to the cuprate superconductivity discovered in the 1980s.[1-8] This has led to the discovery of several new compounds among which is the less toxic FeSe$_x$ [6] that shows zero resistance at 8 K in bulk form which is strongly dependent on Se deficiency and annealing at 400 °C. While the 400 °C anneal is found to reduce the non-superconducting NiAs-type hexagonal phase and increase the PbO-type tetragonal superconducting phase,[6,9] the role of Se deficiency is not well understood. On the other hand the influence of fluorine doping and rare earth substitutions on the superconductivity in the LaO$_{1-x}$F$_x$FeAs compounds have been well understood through, XAS, x-ray emission spectroscopy (XES) and photoemission spectroscopy (PES) studies.[10-12] These investigations show that the Fe 3$d$ states hybridize with the O 2$p$, As 4$p$ and La 6$p$ states which may provide the itinerant charge carriers (electrons) responsible for superconductivity in these compounds. From the XAS and XES measurements supported by density of state (DOS) calculations, it was concluded that these systems exhibit weak to moderate electronic correlations.[10-12] While the substitution leads to electron doping in case of the LaO$_{1-x}$F$_x$FeAs system,[13] how Se deficiency may bring in the mobile carriers in the FeSe$_x$ system leading to superconductivity is not clear. A detailed study of the FeSe$_x$ (x=1-0.8) crystals has been made using XAS Fe and Se K-edge spectra. The results presented here show a lattice distortion and Fe-Se hybridization that are responsible for producing itinerant charge carriers in this system. The lattice distortion was also confirmed by the powder x-ray diffraction (XRD) measurements.

## II. EXPERIMENTAL SECTION

FeSe$_x$ crystals were grown by the high temperature solution method as described earlier.[8,9] Crystals measuring 5 mm x 5 mm x 0.2 mm with (101) plate like habit could be obtained by this method. The grown crystals were characterized for Se deficiency by XRD using a Philips Xpert system and a Joel scanning electron microscope (SEM) coupled with an energy dispersive x-ray spectrometer (EDS) (not presented here). The XAS measurements at the Fe and Se K-edge were carried out on the 17C1 and 01C Wiggler beamlines at the National Synchrotron Radiation Reach Center (NSRRC) in Taiwan, operated at 1.5GeV with a current of 200~240mA. Monochromators with Si (111) crystals were used on both the beam lines with an energy resolution ΔE/E better 2x10$^{-4}$. The Fe and Se K-edge absorption spectra were recorded by the fluorescence yield (FY) mode at room temperature using a Lytel detector.[14] All spectra were normalized to a unity step height in the absorption coefficient from well below to well above the edges. Standard Fe and Se metal foils and oxide powders, SeO$_2$, FeO, Fe$_2$O$_3$ and Fe$_3$O$_4$ were used for energy calibration and also for comparing different electronic valence states. Since surface oxidation was suspected to interfere with the interpretation of the spectra, the FeSe crystals were cleaved *in situ* in vacuum before recording the spectra. Even though we have studied a number of FeSe$_x$ crystals with $x$=1 to 0.8 the results for only three compositions are presented here for comparison and clarity.

## II. RESULTS AND DISCUSSION

### A. microstructure

To facilitate the discussion that follows the building blocks of the tetragonal crystal structure of FeSe the Se-Fe tetrahedra and Fe-Se pyramidal sheets are shown in Fig 1(a) and the electronic structure of individual elements and the FeSe (indicating the hybridization band) are given in Fig 1(b). Fig. 2 shows the XRD patterns of the FeSe$_x$ ($x$=0.9, 0.88 and 0.85) crystals which are found to represent the superconducting α-FeSe phase. The patterns have been fitted to the P4/nmm and indexed in the figure. Weak hexagonal reflections are also seen among these. The main diffraction peak (101) shown expanded in the inset is found to shift to a lower 2θ as $x$ decreases indicating an increase in the lattice parameters. The lattice parameters calculated from these patterns are $a=b=3.771$Å, $c=5.528$Å for $x$=0.9, $a=b=3.775$Å, $c=5.528$Å for $x$=0.88 and $a=b=3.777$Å, $c=5.529$ Å for $x$=0.85. It is observed that the $a=b$ parameter increase though very slightly more than that of the $c$ parameter as $x$ decreases. Thus, the $ab$ plane variation is found to be larger than that of the $c$ axis. These lattice parameters are very close to those reported in literature for Se deficient powders.

**B. Formal charge of Fe and Se in the crystals**

The transition metal Fe K-edge (1s➔4p) XAS spectra are mostly related to the partial density of Fe 4$p$ states at the iron site (Fig. 1(c)). The unoccupied states of the 3$d$ (due to quadruple transition) and 4$sp$ bands are sensitive to the local structure and the type of the nearest neighbors.[15-18] These spectra could therefore be used to obtain information on the changes in the electronic states that may result from changes in the environment of the Fe ions such as reduction in the Se content in the present case. The information so obtained could be used to explain other properties. The Fe K-edge absorption spectra of FeSe$_x$ crystals are presented in Fig. 3 along with the standards Fe, FeO, Fe$_2$O$_3$ and Fe$_3$O$_4$ and are normalized at the photon energy ~100 eV above the absorption edge at E$_0$ = 7,112 eV (the pure Fe K absorption edge energy). The spectra of the crystals appear to be close to the Fe metal foil indicating that the crystals are free from oxidation. This was also confirmed from the recent Fe L-edge and O K-edge spectra measured after cleaving the samples in UHV[‡] (given in supporting data) which show only the Fe 3$d$ band. Thus the investigations presented here on the FeSe system are free from the effects of oxidation ruling out the possibility of the role of oxygen in the superconductivity. as is case in the LaO$_{1-x}$F$_x$FeAs system.

Three prominent features A$_1$, A$_2$ and A$_3$ are observed in these spectra (Fig. 3(a)) of which A$_1$ could be assigned to the 3$d$ unoccupied states originating from the Fe-Fe bonds in metallic iron. The features A$_2$ and A$_3$ represent the unoccupied Fe 4$sp$ states. The rising part of the broad feature A$_2$ (~7,118.8 eV) appears as a broad peak labeled e$_1$ at ~7,116.8 eV, seen well separated in the first derivative plots of these spectra given in Fig. 3(b). It is not seen in those of the reference Fe metal foil or oxide powders and is a part of the Fe 4$sp$ band. The e$_1$ feature appears at an energy between the Fe metal and FeO and therefore has its origin in a different interaction as will be seen. From these first derivative plots, a formal charge of Fe was evaluated in conjunction with the three standards FeO (Fe$^{2+}$), Fe$_2$O$_3$ (Fe$^{3+}$) and Fe$_3$O$_4$ (Fe$^{2.66+}$). In addition we used sine function fitting of the broad feature e$_1$ (7,113.84-7,122.2 eV). By an extrapolation of the energy of FeSe$_{0.88}$ with the standards we obtained a formal charge of ~1.8+ for Fe in these crystals thus establishing the electronic charge of Fe in the covalency (2+). It is also seen that the peak position is not increasing in energy as x is decreased meaning that the effective charge (valence) of Fe does not change with $x$. (We used formal charge as this is not strictly an ionic compound) Thus the possible electronic configurations of Fe in the ground state could be written as 3$d^{6.2}$ or 3$d^6 4s^{0.2}$ indicating a mixture of monovalency (3$d^6 4s^1$ or 3$d^7$) and divalency (3$d^6$).

The Se K-edge spectra of the FeSe$_x$ crystals and Se and SeO$_2$ standards are presented in Fig. 4 (a) and the corresponding first derivative plots are given in Fig. 4(b) to highlight the energy changes in the spectra. The spectra represent mainly Se character without a trace of SeO$_2$, indicating the total absence of oxidation even in the deeper layers of (bulk) FeSe$_x$ crystals. The spectra exhibit two peaks B$_1$ and B$_2$. The B$_1$ feature at photon energy around 12,658 eV is formally assigned to the transitions 1$s$➔4$p$ and shows a slight increase in intensity as well as shift to higher energy as $x$ is decreased. This indicates an increase in the Se 4$p$ unoccupied states. From the first derivative plots a formal charge of ~2.2- is obtained for Se in the $x$=0.88 crystal by interpolation with the energies of the standards Se and SeO$_2$ (as in the case of Fe). This is in agreement with a total charge of 0 when the formal charges of Fe and Se are added (Fe$^{1.8+}$Se$_{0.88}^{2.2-}$). This establishes the electronic charge of Se in the covalency (2-).

**C. Fe 3d-Se 4p hybridization and superconductivity**

The excess negative charge of -0.2 seen on Se may be explained thus: The electronic charge of Fe in the covalent FeSe should be 2+. However we have obtained a formal charge of 1.8+ which means that some electronic charge is returned to Fe due to the Se deficiency. On the other hand the covalent charge of Se should be 2- but we obtain 2.2-, which is beyond the 6 electron occupancy of the 4$p$ state. A hole increase is seen from the Se K-edge spectra as $x$ decreases but no change is observed in the Fe K-edge to suggest a change in valence. The close distance between Se and Fe, might result in an increase in the covalence of Fe-Se bond due to the hybridization of the Se 4$p$ and Fe 3$d$ states as pointed by Yamasaki et al.[19] from their soft x-ray photoemission spectroscopy (XPS) measurements. Yoshida et al.[20] also observe a good correspondence between their DOS calculations and the XPS spectra which show a feature corresponding to the Fe 3$d$-Se 4$p$ hybridization. Theoretical calculation by Subedi et al. [21] also point to this Fe-Se hybridization. From the discussion above, the B$_1$ feature (Fig. 4(a)) therefore represents the Fe 3$d$-Se 4$p$ hybridization band (Fig. 1(c)). The increased electronic charge on Fe seen above is in fact due to itinerant electrons in the Fe-Se hybridization bond and appears as a hole increase in the Se K-edge spectra. From the discussion above it may be concluded that the itinerant carriers are electrons on Fe or in the Fe-Se bond. If it were holes we would have observed a decrease in the effective charge of Se and an increase in that of Fe. This could be confirmed from the Fe L-edge measurements on similar crystals. When we correlate this with the decreasing transition width in the resistance measurements (presented in the supporting data), it could be concluded that the charge carriers responsible for superconductivity are in fact itinerant electrons in a similar manner as the itinerant holes in the case of cuprates. Oxygen annealing in case of YBa$_2$Cu$_3$O$_{6+\delta}$, is found oxidize Cu$^{2+}$ to Cu$^{3+}$ through the hybridization of Cu 3$d$-O 2$p$ states. The Cu$^{3+}$ state has been assigned to the empty state in the Cu-O bonds that is also referred to as the 3$d^9$L ligand state, where a hole is located in the oxygen ions surrounding a 'Cu site' (L, a ligand hole, tentatively label this as 3$d^8$-like).[22] These itinerant holes are responsible for superconductivity. In a similar manner in the case of FeSe, the Se deficiency is bringing a Fe 3$d$-Se 4$p$ hybridization leading to itinerant electrons.

### D. Lattice distortion, structure modulation

It is seen that the intensity of the A$_2$ feature reduces as x is decreased which indicates a lattice distortion that increases with decreasing $x$. In addition the change in the A$_2$ feature is larger than the A$_3$ feature. Since multiple scattering in the XAS from $p$ orbitals could reveal the different orbital orientations and because of the nature of the $p$ orbitals i.e. $p_{xy}$ and $p_z$ the A$_2$ feature could be associated with $p_{xy}$ ($\sigma$) and the A$_3$ to $p_z$ ($\pi$) orientations. This leads us to a speculation of a larger distortion in the $ab$ plane (Fe-Fe distance) compared to the $c$ axis. This is also seen from the XRD measurements (Fig. 2) where the change in the $a=b$ parameter is larger than that in the $c$ parameter (Fe-Se distance). Therefore, the Fe orbital structure changes from 4$p$ to a varying (modulating) coordination as $x$ is decreased. The broad feature B$_2$, at ~ 20 eV above the Se K-edge appears at the same energy in all the FeSe$_x$ spectra and is not affected by the Se deficiency and is assigned to the multiple scattering from the symmetrical Se 4$p$ states in the coordination sphere that are correlated to the local structure of the Se ions.[15] This is consistent with the XRD result where the c parameter is nearly unchanged. This becomes clear by looking at Fig. 1 where Se is seen at the tip of the Fe-Se pyramid.

Since Se is located at the apex of the tetrahedral pyramid chain in the FeSe lattice (as shown in Fig. 1), the removal of a Se ion with formal negative charge (-2) from the lattice would result in a Se vacancy with an effective positive charge and would cause repulsion to the surrounding positively charged Fe atoms. This would lead to greater interaction between the Fe-ions in the $ab$ plane and a lattice distortion with a structure modulation. This is consistent with the distortion in the $ab$ plane discussed above. In addition the Fe atoms around the vacancy may act like magnetic clusters as pointed out by Lee et al..[23]

### IV. CONCLUSION

In conclusion a lattice distortion observed in the XAS Fe K-edge spectra of Se deficient FeSe$_x$ crystals is concluded to produce itinerant electrons in the Fe-Se hybridization bond seen in the Se K-edge spectra. The XRD measurements confirm this lattice distortion that increases with Se deficiency. The charge balance considerations from Se deficiency also result in itinerant electrons (in the Fe-Se hybridization bond). The itinerant electrons are found to reduce the width of the resistive (superconducting) transition (in supporting data). The symmetry of the Fe in the $ab$ plane changes from the 4 $p$ orbital to modulating (varying) coordination geometry.

### Acknowledgments

This work was supported by the National Science Council of R. O. C. under contracts NSC96-2112-M-001-026-MY3 and NSC96-2811-M-001-071. One of the authors (SMR) is grateful to the NSC for financial support. The experimental support from NSRRC is gratefully acknowledged.

**Figure captions:**

FIG. 1 (color online) (a) Illustration of the crystal structure of tetragonal FeSe, the blue (small balls) and red (large balls) denote Fe and Se, respectively; the pyramid chain and tetrahedral arrangements are marked in color in the unit cell. The hybridization is shown in two color bonds; (b) the local symmetry of Se atom (in pyramid chain) and Se atom (in tetrahedral geometry) shown separately; (c) energy level diagram of the FeSe system along with the individual elements. The hybridization and unoccupied states in the FeSe are highlighted by a circle.

FIG. 2 (color online) Powder XRD patterns of FeSe$_x$ crystals with $x=$ (i) 0.85, (ii) 0.88 and (iii) 0.9. The patterns are fitted to the P*4/nmm* space group and indexed. The hexagonal phase reflections are marked with a prefix H.

FIG. 3 (color online) (a) Fe K-edge (1s→4p) absorption spectra of FeSe$_x$ with different Se content and (b) the first derivative plots of the same spectra. The spectra of the standards - Fe-metal foil, FeO, Fe$_2$O$_3$ and Fe$_3$O$_4$ are also given alongside the sample spectra.

FIG. 4 (color online) (a) The Se K-edge (1*s*→4*p*) absorption spectra FeSe$_x$ with different Se content along with the standards Se metal and SeO$_2$ and (b) the first derivative plots of the same spectra.

**ADDITIONAL INFORMATION**

R-T plots of the crystals with x=0.95 to 0.8 are given in Fig. A1. It is seen that the resistance normalized at 280 K (R/R280) in all the plots exhibits similar behavior, going from a metallic to superconducting, with the anomaly at 100 K [6,8] nearly unchanged. The expanded view of the Superconducting transition region given in the inset shows that the width of the transition decreases initially at x=0.9 and increases thereafter. The $T_0$ also shows a maximum at =.0.9 and increases thereafter. This may be a result of increase in carrier concentration at 0.9 as in the case of cuprates [9] or some other phenomenon like the magnetic phases getting removed at this composition that facilitate the onset of superconductivity. [24]

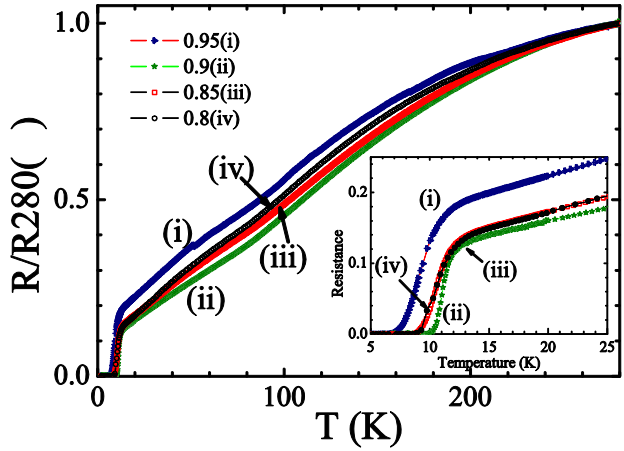

FIG. A$_1$   The R-T plots of FeSe$x$ (i) x=0.95, (ii) 0.9, (iii) 0.85 and (iv) 0.8.  The anomaly at 100 K is not changed with $x$ (unlike the LaO$_{1-x}$F$_x$FeAs system). The inset shows that the transition width is smallest at $x$=0.9 (close to 0.88) as pointed decreasing with decreasing $x$ initially.

**SUPPORTING DATA**

‡*Measurements made at beamline 7 of the Advanced Light source, Berkley, CA, USA.*

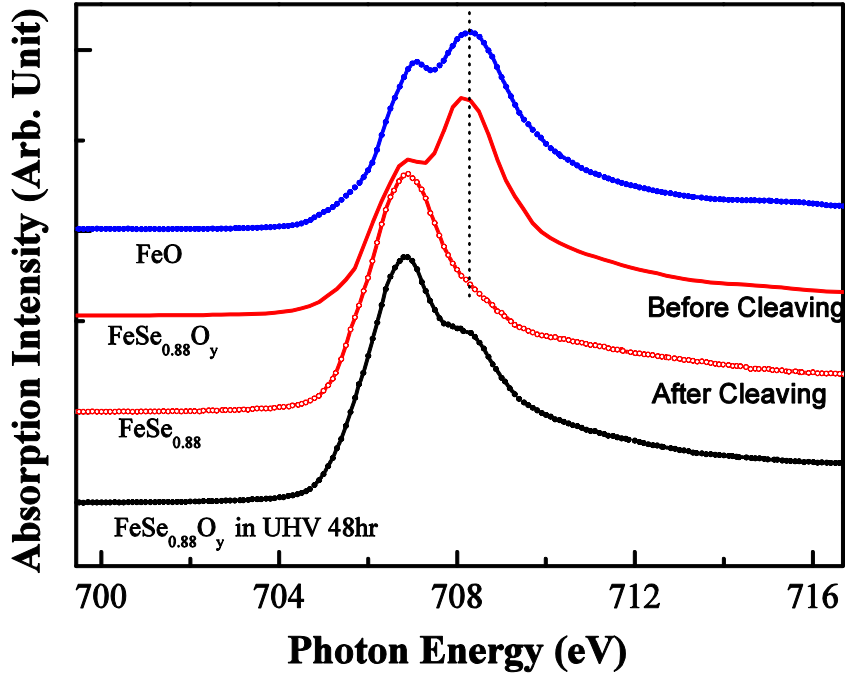

FIG. A$_2$  XAS Fe L$_3$-edge spectra of FeSe$_{0.88}$ crystal before and after cleaving *in situ* in vacuum.

The Fe L$_3$-edge (2$p$2/3→3$d$ transition) spectra recorded on the FeSe crystal that was separated in dry condition without washing in water to remobe KCl flux, before and after it was cleaved in UHV are shown in this figure. The oxidation peak observed in the crystal before cleaving is not observed after cleaving. The pure FeSe$_x$ spectral white line spectra measured on crystals with different $x$ value i. e. with different Se concentration did not show much difference. We have presented measurements only on the FeSe$_{0.88}$ for clarity as well as much magnetic and transport data is available on crystals with this composition. The spectral line shapes of the cleaved crystals resemble those of ion metal suggesting a strong covalent Fe-Fe bounding in the interlayer of the Fe plane. More detail about the FeSe electronic structure of the 3$d$ states obtained by XAS measurements and resonant inelastic x-ray scattering (RIXS) at the Fe L$_{2,3}$ edges will be discussed separately.[25] Our observations are support the results of W. L. Yang *et al.* [12] on Fe-pnictides 1111 and 122 systems and K. W. Lee *et al.* [23] using first principles methods to study FeSe system. These results are consistant with the Fe K-edge spectra presented above as well as the EDS analysis.

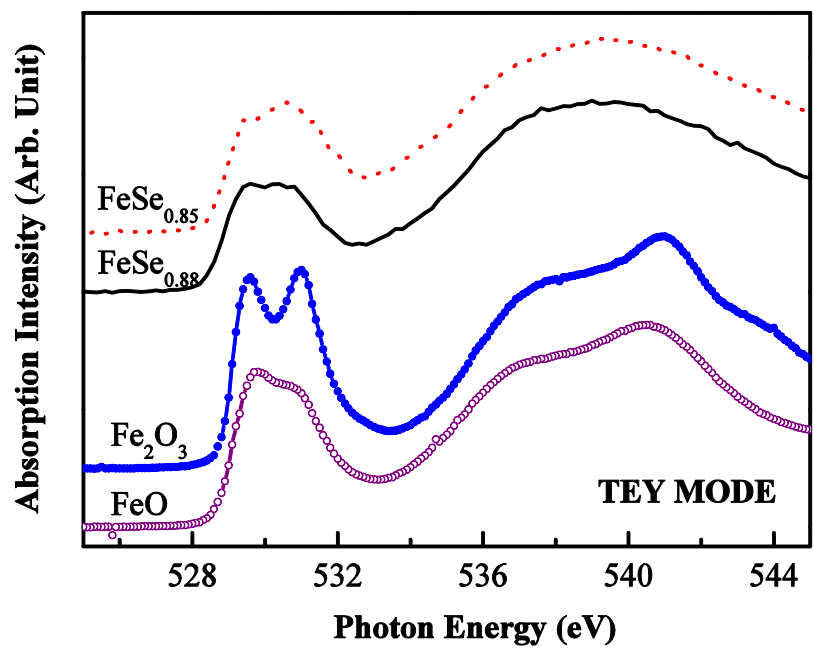

FIG. A$_3$ O K-edge XANES spectra of a series of FeSe$_x$ crystals and the standards Fe$_2$O$_3$ and FeO.

The O K-edge XANES of spectra are closely related to the partial density of O 2$p$ unoccupied state in the transition metal oxides and are sensertive to the oxidation states as seen in the Figure. The pre-edge feature around 528-533 eV diplays the Fe metal $d$ and oxygen $p$ hybridiztion band at the oxygen site and the main feature at highest position ~540eV corresponds to the $s$-$p$ band of the oxygen. According to the absorption data collected in total electron yield (TEY) mode in the solft x-ray region, where the probing depth is only a few angstroms, the surfaces were found to be FeO like as seen from the figure above. The spectra recorded on the FeSe$_x$ samples indicate the presence of a small amount of oxygen but not the formation of a stable compound like FeO or Fe$_2$O$_3$. These results are consistant with the Fe K-edge spectra presented above as well as the EDS analysis (not presented here). These layer structured crystals and could be cleaved in the ultra high vacuum (UHV) chamber. We therefore took a crystal that was not washed in water to remove KCl but removed carefully from the KCl in dry condition and cleaved in the UHV chamber. The result was spectacular and confirmed the earlier doubt that the washing out the KCl in water might cause the surface oxidation of the crystals.

# Figures

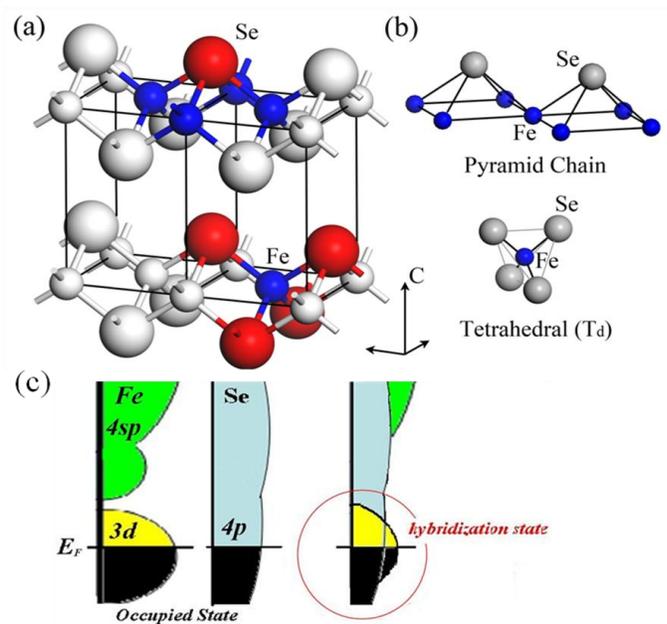

FIG. 1

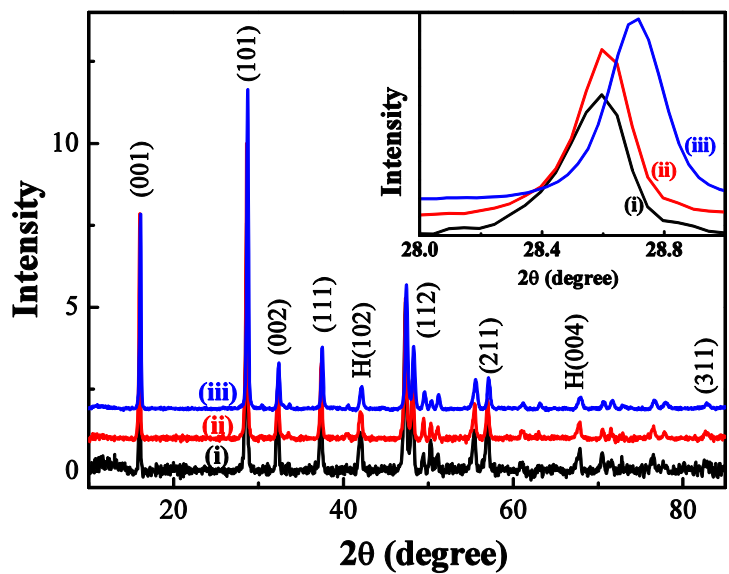

FIG. 2

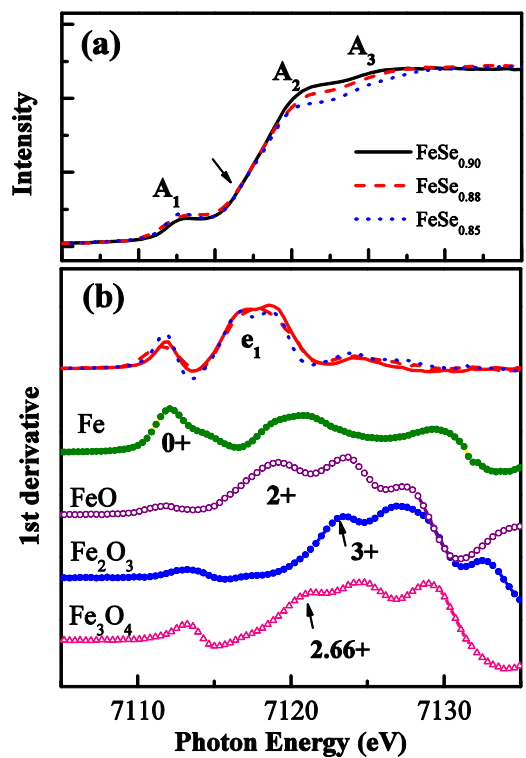

FIG. 3

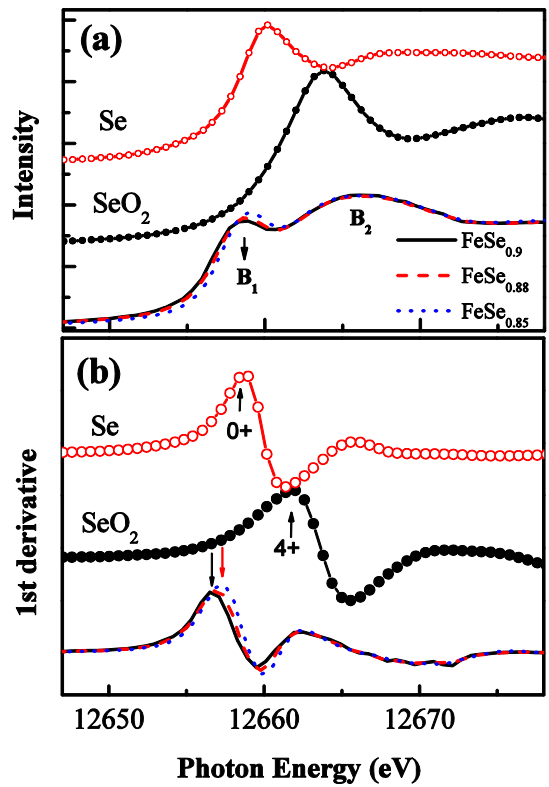

FIG. 4